\documentclass[aps,superscriptaddress,twocolumn,floatfix,longbibliography]{revtex4-1}
\usepackage[utf8]{inputenc}
\usepackage{physics}
\usepackage{siunitx}
\usepackage{subfigure}
\usepackage{hyperref}
\usepackage{blindtext}
\usepackage{amsmath}
\usepackage{mathrsfs}
\usepackage{latexsym}
\usepackage{amsfonts}
\usepackage{amssymb,amscd}
\usepackage[all]{xy}
\usepackage{amsthm}
\usepackage{fancyhdr}
\usepackage{fancybox}
\usepackage{xcolor}
\usepackage{verbatim}
\usepackage{makeidx}
\usepackage{indentfirst}
\usepackage[english]{babel}
\usepackage{graphicx}
\usepackage{listings}
\usepackage{multirow}
\usepackage{framed}
\usepackage{bbold}
\usepackage{cases}
\usepackage{ulem}
\usepackage{enumitem}
\usepackage{lipsum}
\usepackage[portrait, margin=1in]{geometry}

\newcommand{\rdbkt}[1]{\left( #1 \right)}
\newcommand{\sqbkt}[1]{\left[ #1 \right]}
\newcommand{\crbkt}[1]{\left\{ #1 \right\}}

\renewcommand{\rm}[1]{\mathrm{#1}}

\newcommand{\Ms}{M_\mathrm{s}}

\begin{document}

% Here are two options for the title:
%\title{\bl{Evidence for Chiral Quantized Spin Waves in Nanowires with Antisymmetric Exchange Interactions Using Brillouin Light Scattering Spectroscopy}}

\title{Brillouin Light Scattering from Quantized Spin Waves in Nanowires with Antisymmetric Exchange Interactions}

\author{Jun-Wen Xu}
\affiliation{Center for Quantum Phenomena, Department of Physics, New York University, NY 10003, USA}
\author{Grant A. Riley}
\affiliation{Quantum Electromagnetics Division, National Institute of Standards and Technology, Boulder, CO 80305, USA}
\affiliation{Center for Memory and Recording Research, University of California – San Diego, La Jolla, CA 92093, USA}
\author{Justin M. Shaw}
\affiliation{Quantum Electromagnetics Division, National Institute of Standards and Technology, Boulder, CO 80305, USA}
\author{Hans T. Nembach}
\affiliation{Quantum Electromagnetics Division, National Institute of Standards and Technology, Boulder, CO 80305, USA}
\affiliation{Department of Physics, University of Colorado, Boulder, CO 80309, USA}
\author{Andrew D. Kent}
\email[]{andy.kent@nyu.edu}
\affiliation{Center for Quantum Phenomena, Department of Physics, New York University, NY 10003, USA}

\begin{abstract}
Antisymmetric exchange interactions lead to non-reciprocal spin-wave propagation. As a result, spin waves confined in a nanostructure are not standing waves; 
they have a time-dependent phase, because counter-propagating waves of the same frequency have different wavelengths. 
We report on a Brillouin light scattering (BLS) study of confined spin waves in Co/Pt nanowires with strong Dzyaloshinskii–Moriya interactions (DMI).
Spin-wave quantization in narrow ($\lesssim \SI{200}{nm}$ width) wires dramatically reduces the frequency shift between BLS Stokes and anti-Stokes lines associated with the scattering of light incident transverse to the nanowires.
In contrast, the BLS frequency shift associated with the scattering of spin waves propagating along the nanowire length is independent of nanowire width.
A model that considers the chiral nature of modes captures this physics and predicts a dramatic reduction in frequency shift of light scattered from higher energy spin waves in narrow wires, which is confirmed by our experiments.
\end{abstract}
\maketitle

Antisymmetric exchange interactions fundamentally change the nature of spin-wave excitations and ground-state spin configurations.
These interactions were first considered by Dzyaloshinskii~\cite{dzyaloshinsky1958thermodynamic} and Moriya~\cite{moriya1960anisotropic} to explain the origin of the small magnetic moment in several uncompensated antiferromagnetic materials, which are now known as Dzyaloshinskii–Moriya interactions (DMI).
In contrast to Heisenberg exchange interactions, which lead to the collinear alignment of neighboring spins, DMI results in spin canting and chiral spin textures, such as magnetic skyrmions~\cite{fert2017magnetic,buttner2017field,caretta2018fast}.
These topological magnetic objects are of intense interest in basic physics~\cite{roessler2006spontaneous,uchida2006real} and for possible racetrack memory devices~\cite{parkin2008magnetic,tomasello2014strategy,parkin2015memory}.

A key characteristic of materials with DMI is non-reciprocal spin-wave propagation, with different wavevectors and characteristics for left and right propagating spin waves~\cite{bogdanov2001chiral,cortes2013influence,moon2013spin,di2015asymmetric,lee2016all,zingsem2019unusual,wang2020chiral,lucassen2020extraction}, i.e., the spin-wave dispersion is no longer symmetric about zero wavevector.
The consequences of DMI are most directly observed in Brillouin light scattering (BLS) experiments in which photons create and annihilate spin waves with wavevectors collinear with the incident light.
Since the wavevectors for these two processes have opposite signs, the frequency shift of the light is a direct measure of the non-reciprocal nature of the exchange interactions.
For this reason, BLS is now a technique of choice for characterizing DMI~\cite{nembach2015linear}.

The presence of DMI leads to interesting new physics in the case of confined spin waves~\cite{zingsem2019unusual}.
Here, interference between counter-propagating spin waves cannot lead to standing waves, as left and right propagating waves at the same energy have different wavelengths.
The quantized spin-wave modes also do not have space-inversion symmetry. 
The consequences of spin-wave quantization effects in the presence of DMI have not, to our knowledge, been studied experimentally.

In this article we present a Brillouin light scattering study of spin waves in nanowires with strong DMI.
A dramatic reduction of the frequency shift between counter-propagating confined spin-wave modes occurs as the wire width is reduced.
This is a direct consequence of the unusual non-standing-wave nature of the quantized spin-wave modes in the presence of chiral magnetic interactions. 

\begin{figure}[t]
    \centering
    \includegraphics[width=0.48\textwidth]{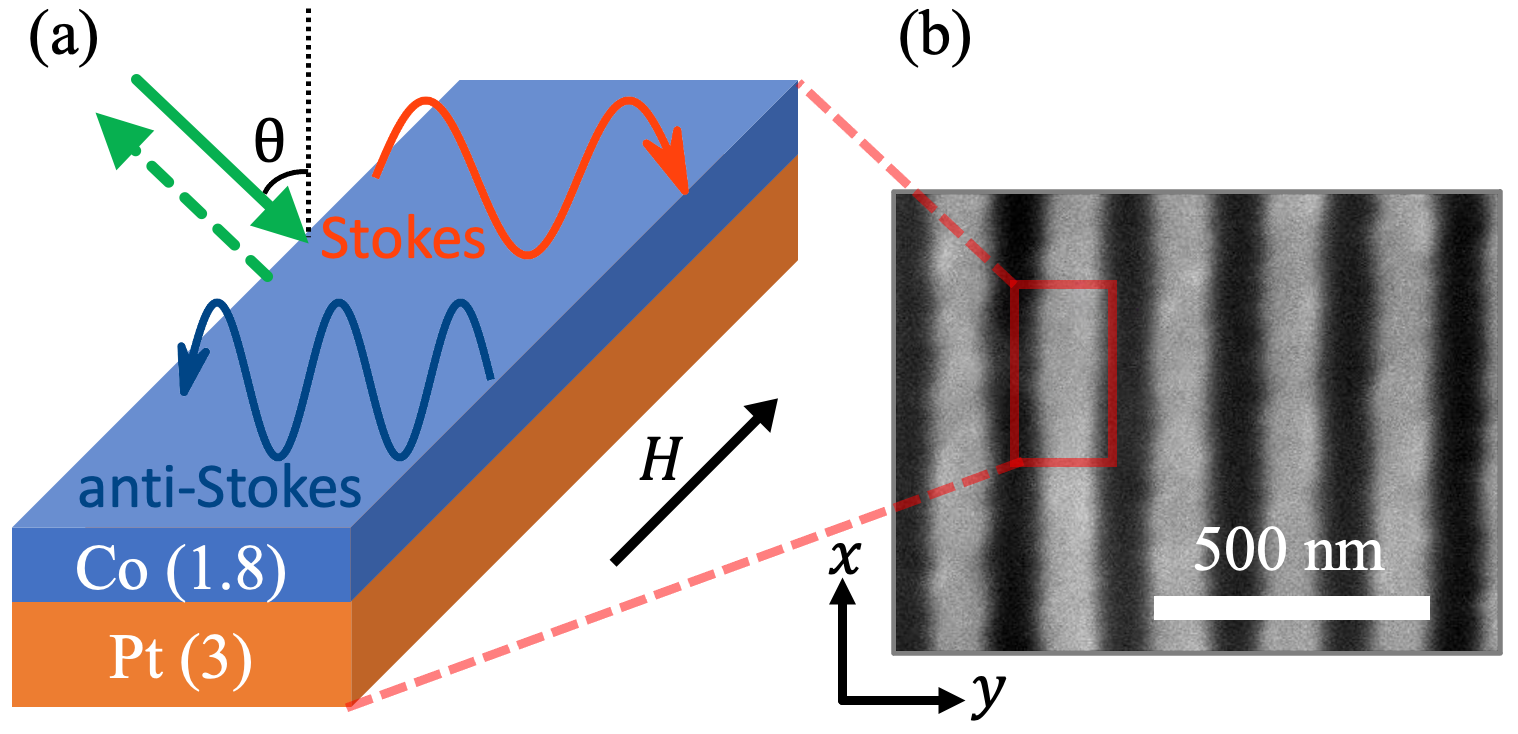}
    \caption{(a) Schematic of spin waves in a confined geometry with chiral magnetic interactions.
    Left and right propagating spin waves of equal energy have different wavevectors, as indicated by the blue and orange colored waves.
    In BLS, light with an angle of incidence $\theta$, perpendicular to the applied field, $H$, is backscattered by spin waves, as indicated by the green arrows.
    In a Stokes process, magnons with wave vectors moving away from the incident light are created, while in an anti-Stokes process, magnons of the opposite wave vectors are annihilated.
    (b) An SEM image of the nanowire array consisting of $\SI{100}{nm}$ width Co/Pt nanowires with $\SI{100}{nm}$ spacing. The nanowires are aligned in the $x$ direction and their width is in the $y$ direction.}
    \label{fig_BLS_SEM}
\end{figure}

The basics physics is illustrated in Fig.~\ref{fig_BLS_SEM}(a).
In a nanostructure with chiral interactions, spin waves of the same frequency propagating in opposite directions have different wavevector magnitudes, illustrated schematically by the orange and blue curves.
The resulting interference pattern produced by these counter-propagating waves is thus not either symmetric or antisymmetric about the midplane of the nanowire, as is the case for usual standing waves~\cite{zingsem2019unusual}.

BLS can be used to determine the frequency shift associated with the inelastic scattering of light by spin waves, also known as magnons, i.e., a frequency shift caused by photon-magnon interactions.
BLS is a powerful method to measure DMI in thin films and nanostructures~\cite{nembach2015linear,di2015direct,stashkevich2015experimental,quessab2020tuning}.
This is because the frequency shift between counter-propagating spin waves is a direct consequence and measure of the strength of the chiral magnetic interactions.
As shown in Fig.~\ref{fig_BLS_SEM}(a), light is incident at an angle to the film normal and a magnetic field is applied in the film plane perpendicular to the light's plane of incidence, a configuration known as the Damon-Eshbach geometry.
The light backscattered from the sample is collected and analyzed.
When the scattering process creates a magnon, the backscattered photon's frequency decreases (a Stokes process), whereas when a magnon is annihilated in the scattering process the frequency of the photon increases (an anti-Stokes process)~\cite{gubbiotti2010brillouin}.
By energy conservation, the shift of the photon frequency is the frequency of the excited or annihilated spin wave.
The magnon momentum $q_m$ is related to the angle of incidence of the light $\theta$; momentum conservation gives, $q_\rm{m} = \rdbkt{4 \pi/ \lambda} \sin \theta$, where $\lambda$ is the wavelength of the light.

We conducted BLS on ferromagnetic nanowire arrays fabricated from Ta(4)/Pt(3)/ Co(1.8)/Al(2)/Pt(3) thin films on oxidized silicon wafers, deposited using dc magnetron sputtering, with the numbers being the layer thicknesses in nanometers.
The Pt/Co interface has been shown to induce a large DMI~\cite{belmeguenai2015interfacial}.
The Al layer decouples the Co layer and the Pt cap layer, which also protects the film from oxidation.
Electron-beam lithography followed by Ar ion milling was used to define nanowire arrays and the width/spacing was varied from $\SI{100}{}$ to $\SI{400}{nm}$.
A scanning electron microscope (SEM) image of $\SI{100}{nm}$ width nanowires is shown in Fig~\ref{fig_BLS_SEM}(b).
Arrays are needed to have a sufficient filling factor for the BLS laser spot size \footnote{In principal micro-BLS could be used to study a single nanowire. However, its diffraction-limited resolution is of the order of \SI{300}{nm}, which is not sufficient for our study. For spatially resolved images, 
a wire width of a few micrometers would be needed to have multiple data points across the standing wave. But spin-wave quantization effects for a micron width wire would be negligible.}.
\begin{figure}[t]
    \centering
    \includegraphics[width=0.48\textwidth]{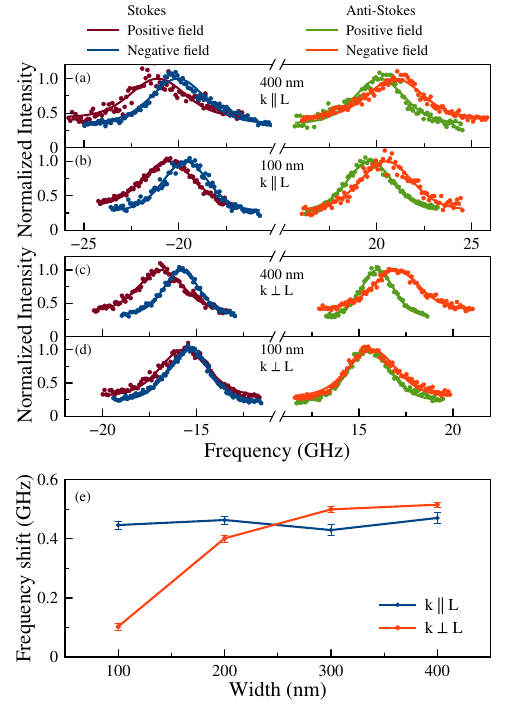}
    \caption{(a)-(d) BLS spectra for Stokes and anti-Stokes processes for both positive and negative fields on nanowire arrays with different widths.
    (e) The absolute value of the frequency shift on nanowire arrays with different widths in the longitudinal ($k \parallel L$) and transverse ($k \perp L$) scattering geometries. In the longitudinal geometry the frequency shift is independent of nanowire width. Whereas in the transverse geometry the frequency shift decreases significantly in narrow wires.
    The magnetic fields applied $\mu_0 H$ for longitudinal and transverse scattering geometries are $\SI{0.784}{T}$ and $\SI{0.578}{T}$, respectively.}
    \label{Fig:BLS_shift}
\end{figure}

Two different scattering geometries were used: a transverse geometry in which the incident light is perpendicular to the nanowire, as illustrated in Fig.~\ref{fig_BLS_SEM}(a), and a longitudinal geometry, where the incident light is parallel to the nanowire.
BLS experiments were conducted with light of wavelength $\SI{532}{nm}$ at an incident angle of 45$^\circ$, giving a momentum transfer $q_m=16.7$ $\mu$m$^{-1}$ (see Appendix Sec.~\ref{Sec:Appendix_Angles} for results at different angles of incidence). 
Figures~\ref{Fig:BLS_shift}(a)-(d) show the spectra of $\SI{400}{}$ and $\SI{100}{nm}$ nanowire arrays in these two geometries.
The spectra indicate the Stokes (negative frequencies) and anti-Stokes (positive frequencies) lines for both field polarities.
The scattering intensities are fit to find the peak positions and thus the frequency shift between the Stokes and anti-Stokes lines.

As noted, the frequency shift is a direct consequence of the spin-wave dispersion being asymmetric with respect to wave vector inversion ($k\rightarrow -k$), i.e.,:
\begin{equation}
    f_k=f_0+\frac{\gamma pDk}{\pi M_\rm{s}}, 
    \label{Eq:FM}
\end{equation}
where $f_0$ is the spin-wave frequency in the absence of the DMI, $\gamma$ is the gyromagnetic ratio, $M_\rm{s}$ is the saturation magnetization, $k$ is the spin-wave vector, $p = \pm 1$ indicates the magnetization polarity with respect to the $x$ direction in the scattering geometry shown in Fig.~\ref{fig_BLS_SEM}, and $D$ is the interfacial DMI.
The frequency shift as a function of wire width in the longitudinal and transverse geometries are shown in Fig.~\ref{Fig:BLS_shift}(e).
In the longitudinal geometry, where the spin waves propagate along the nanowire, the frequency shift is independent of wire width.
This is consistent with Eq.~\ref{Eq:FM} and enables determination of the interfacial DMI from the frequency shift, with the known magnetization and gyromagnetic ratio~\cite{nembach2015linear}.
The latter are determined using magnetometry and ferromagnetic resonance spectroscopy to be $M_\rm{s} = \SI{9.39 \pm0.01 e5}{A/m}$ and $\gamma/2 \pi = \SI{30.3 \pm 0.1}{GHz/T}$, respectively.
We thus find the interfacial DMI to be $D = \SI{4.79 \pm 0.07 e-4}{J/m^2}$. (The Appendix Sec.~\ref{Sec:Appendix_Angles} includes magnetic measurements and BLS results at different angles of incidence.)

However, in the transverse geometry, the frequency shift between the Stokes and anti-Stokes peak positions strongly depends on the nanowire width (Fig.~\ref{Fig:BLS_shift}(c-d)); it decreases by more than a factor of 4 as the wire width varies from $\SI{400}{nm}$ to $\SI{100}{nm}$ (Fig.~\ref{Fig:BLS_shift}(e)).
This large reduction in the frequency shift cannot reflect changes in the DMI, as the DMI interaction is local; it is associated with exchange interactions and spin-orbit coupling on neighboring atoms at the Co/Pt interface~\cite{kim2018correlation}. 
Further, no changes in the spin-wave frequency shift were seen in the longitudinal scattering geometry, which would be affected if there were changes in the magnetic characteristics of the nanowires as their width is reduced.

Instead, we show that the reduction in the BLS frequency shift in the transverse geometry is a direct consequence of the unusual nature of the quantized spin-wave modes in the presence of chiral magnetic interactions~\cite{zingsem2019unusual}.
We consider only magnetic interactions within an individual nanowire, as interwire dipolar interactions are negligible compared to the intrawire exchange and DMI interactions~\cite{gubbiotti2005magnetostatic}.
To illustrate the essential physics, we consider the spin-wave dispersion relation~\cite{kalinikos1986theory,jorzick1999brillouin,moon2013spin}:
\begin{widetext}
\begin{equation}
    f(k) = \frac{\gamma \mu_0}{2 \pi} \left[\sqrt{\left(H + \frac{2 A}{\mu_0\Ms} k^2 \right) \left( H + \frac{2 A}{\mu_0\Ms} k^2  + M_\mathrm{eff} \right)} + \frac{2 p D k}{\mu_0\Ms} \right],
    \label{eq_dispersion}
\end{equation}
\end{widetext}
where $\mu_0$ is the permeability of free space, $H$ is the applied field magnitude, $A$ is the exchange constant and $M_\mathrm{eff}$ is the effective magnetization, the demagnetization field minus the perpendicular anisotropy field associated with the Co/Pt interface $K_p$, $M_\mathrm{eff}$=$M_\rm{s}- 2K_p/(\mu_0 M_\rm{s})$; we do not include the demagnetization factors in the nanowire width dimension as these do not come into the analysis of the BLS results (see Appendix, Sec.~\ref{Sec:DM}).
$A = \SI{2.27 \pm 0.01 e-11}{J/m}$ and $M_\mathrm{eff}= \SI{-7.35 \pm 0.08 e4}{A/m}$ are determined from magnetization measurements and ferromagnetic resonance spectroscopy as discussed in the Appendix.

\begin{figure}[t]
    \centering
    \includegraphics[width=0.48\textwidth]{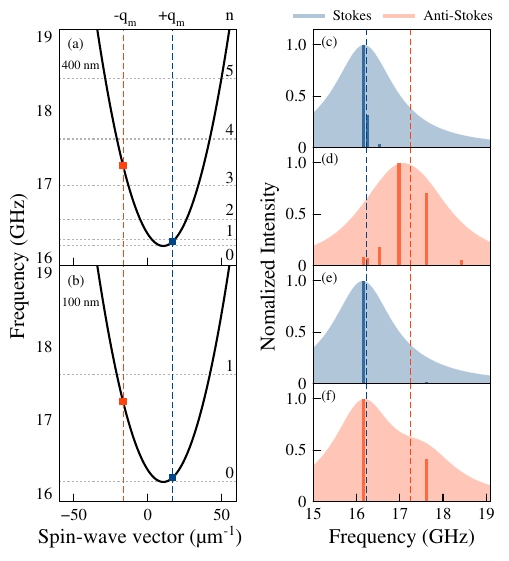}
    \caption{Spin-wave dispersion for (a) $\SI{400}{nm}$ and (b) $\SI{100}{nm}$ nanowires based on Eq.~\ref{eq_dispersion} for negative field polarity($p=-1$).
    The vertical lines indicate the transferred wave vector $q_\rm{m}$ for both Stokes and anti-Stokes processes.
    The blue and orange squares at the intersection of $\pm q_\rm{m}$ with the dispersion curve are the BLS peak positions for Stokes and anti-Stokes process in the continuum limit.
    The gray horizontal lines indicate the frequencies of quantized spin waves, with their indices $n$ labeled on the right side.
    (c-f) Bar graphs of the BLS light scattering intensities for the quantized modes $I_n$ in Eq.~\ref{Eq:Intensity} for (c-d) $\SI{400}{nm}$ and (e-f) $\SI{100}{nm}$ nanowires, respectively.
    The blue figures are Stokes processes, and the orange figures are anti-Stokes processes.
    The blue and orange vertical dashed lines indicate the frequencies in a continuous ($d\rightarrow \infty$) limit as shown as squares in (a,b).
    Spin wave quantization in the narrowest nanowire leads to the scattering intensity being largest for the lowest frequency $n=0$ mode, the most uniform mode, which leads to a reduced BLS frequency shift.
    The shaded colors show the spectra including the finite spin lifetime.}
    \label{fig_analysis}
\end{figure}

In the $y$ (the nanowire width) direction the spin-wave modes are confined and have quantized energies.
At fixed frequency (or energy) the modes can be written as resulting from an interference pattern between left and right propagating spin waves, $\tilde{m}(y,t)=e^{-i2\pi f_k t}m(y)$, with $m(y)$, the $y$-component of the magnetization, given by:
\begin{equation}
    m(y)=\frac{m_0}{2}\sqbkt{e^{i k_1(y+d/2)}+e^{i k_2(y+d/2)}} \, ,
    \label{eq_m}
\end{equation}
where $m_0$ is the oscillation amplitude and $y \in \sqbkt{-d/2, d/2}$; the nanowire has a width $d$ and spans from $-d/2$ to $d/2$ in the $y$ direction.
$k_1$ and $k_2$ are spin-wave vectors corresponding to spin waves propagating to the right and left (i.e., the sign of the wavevector is included in $k_1$ and $k_2$) with the same frequency, $f_k$, and thus energy.
In the absence of DMI, the dispersion $f\rdbkt{k}$ is an even function of $k$, so $k_1 = -k_2$.
With DMI the symmetry between counter-propagating spin-wave vectors is broken, causing $k_1 \neq -k_2$.
Eq.~\ref{eq_m} can be written as:
\begin{eqnarray}
        m\rdbkt{y} =& \frac{m_0}{2} \sqbkt{e^{i k_{1} \rdbkt{y + d/2}} + e^{i k_{2} \rdbkt{y + d/2}}}\\
        =& m_0 \exp \sqbkt{\frac{i \rdbkt{k_{1} + k_{2}}}{2} \rdbkt{y + d/2}}\\
        &\times \cos \sqbkt{k_n \rdbkt{y + d/2}} \, ,
    \label{Eq:Notstanding}
\end{eqnarray}
where,
\begin{equation}
    k_n=(k_{1}-k_{2})/2=n\pi/d, \;\;n = 0, 1, 2 ,\cdots
    \label{Eq:kn}
\end{equation}
sets the quantization condition.
Here unpinned boundary conditions are assumed~\cite{jorzick1999brillouin}.  
This form has an envelope (set by $k_n$) with a beat structure (given by $k_{1}+k_{2}$)~\cite{zingsem2019unusual}.
The gray horizontal lines in Fig.~\ref{fig_analysis}(a,b) indicate the quantized spin-wave frequencies for $\SI{400}{nm}$ and $\SI{100}{nm}$ width nanowires, respectively.

We now consider the scattering of light from these quantized modes.
BLS is associated with magneto-optic effects in which light can be considered to be Bragg reflected from a phase grating created by spin waves.
The differential light-scattering cross section for in-plane momentum transfer $q$ is proportional to $I(q) = |m_q/m_0|^2$ where~\cite{jorzick1999brillouin}:
\begin{widetext}
\begin{equation}
    \frac{m_q}{m_0} = \frac{1}{m_0} \int_{-d/2}^{d/2} m(y) e^{- i q y} \dd{y} = \frac{d}{2} \crbkt{ e^{i k_1 d/2}\rm{sinc} \sqbkt{\frac{1}{2} \rdbkt{k_{1} - q}d} + e^{i k_2 d/2} \rm{sinc} \sqbkt{\frac{1}{2} \rdbkt{k_{2} - q}d}}\, ,
\end{equation}
\end{widetext}
where $\rm{sinc} \, x \equiv \sin x /x$.
For quantized spin waves described by Eq.~\ref{Eq:Notstanding} the normalized BLS intensity is given by:
\begin{widetext}
\begin{equation}
    I_n(q, d) = \frac{d^2}{4} \crbkt{\rm{sinc} \sqbkt{\frac{1}{2} \rdbkt{k_{1} - q}d} + \rdbkt{-1}^n \rm{sinc} \sqbkt{\frac{1}{2} \rdbkt{k_{2} - q}d}}^2 \, ,
    \label{Eq:Intensity}
\end{equation}
\end{widetext}
where $k_{1}$ and $k_{2}$ are set by the quantization condition (Eq.~\ref{Eq:kn}).
Eq.~\ref{Eq:Intensity} includes terms of the form of sinc functions.
This means that if $k$ were a continuous function (not quantized), the intensity would be maximum when $k_{1} = q_\rm{m}$ for the Stokes process and $k_{2} = -q_\rm{m}$ for the anti-Stokes process, indicated by the blue and orange squares in Fig.~\ref{fig_analysis}(a,b).

Figure~\ref{fig_analysis}(c-f) shows the normalized intensities calculated for each quantized mode $n$ using Eq.~\ref{Eq:Intensity} for $\SI{400}{nm}$ and $\SI{100}{nm}$ nanowires both for Stokes and anti-Stokes processes.
The results are shown as bar graphs.
For the $\SI{400}{nm}$ sample, the maximum intensities occur for the $n=0$ mode for the Stokes process and $n=3$ for the anti-Stokes process.
Their frequency difference is very close to that expected in the continuum limit, indicated by the dashed blue and orange vertical lines.
However, for the $\SI{100}{nm}$ nanowire, the maximum scattering intensities for both the Stokes and the anti-Stokes processes are associated with the $n=0$ mode.
As a result, the frequency difference between the maxima is zero, i.e., in both cases maximum scattering intensity is associated with the lowest frequency and the most spatially uniform mode. 

\begin{figure}[t]
    \centering
    \includegraphics[width=0.48\textwidth]{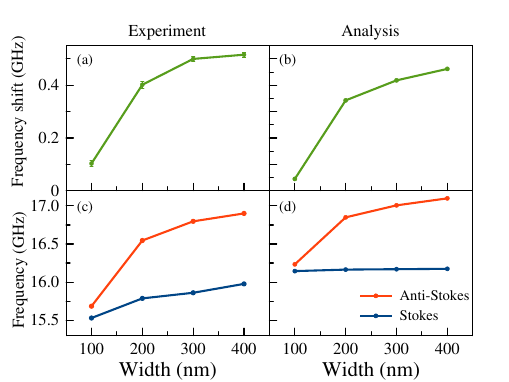}
    \caption{(a) Measured BLS frequency shift for different nanowire widths.
    (b) Frequency shift determined from the quantized spin-wave model.
    (c) The measured BLS Stokes and anti-Stokes peak frequencies for negative field polarity. 
    (d) Model peak frequencies for both Stokes and anti-Stokes processes; the model predicts a stronger decrease in the anti-Stokes frequency which is observed in the experiment.}
    \label{fig_compare}
\end{figure}

This is the basic physics: Spin-wave quantization leads to the BLS light scattering from the narrowest nanowire being dominated by the lowest frequency and the most uniform mode, which is least affected by DMI, because of its small wavevector.
To make a more quantitative comparison between the model and experiment, we consider the lifetime of the modes by convoluting the intensities associated with the quantized modes with a Lorentzian determined by the mode lifetime, set by the damping (see the Appendix Sec.~\ref{Sec:Appendix_Lifetime}).
The shaded colors in Fig.~\ref{fig_analysis}(c-f) are the resulting intensity profiles.
After considering the spin state lifetimes, we determine and plot the resulting peak frequency shift versus wire width.
This is shown in Fig.~\ref{fig_compare}(b) next to the experimental results in Fig.~\ref{fig_compare}(a).
The model captures the experimental trends well.

It is clear that damping broadens the spectra so that it is not possible to observe distinct peaks associated with the quantized modes. This is a consequence of the experimental requirement of having a large DMI; a large DMI requires the ferromagnetic layer to be thin and spin pumping in such layers increases the damping significantly~\cite{boone2015spin}. This would be the case independent of the intrinsic damping of the ferromagnetic layers (which can be very low in transition metal alloys~\cite{Shaw2016}). Increased damping is also founded in other studies of nanostructured samples~\cite{shaw2009spin}.
Furthermore, the broad peaks may also partly be a result of inhomogeneous broadening, as we are measuring an array of nanowires, instead of a single nanowire.

The model further predicts that the decrease of the BLS frequency shift is mainly associated with a reduction in the frequency of the anti-Stokes peak.
The experimental results are shown in Fig.~\ref{fig_compare}(c) and the model results are in Fig.~\ref{fig_compare}(d).
The anti-Stokes frequency is indeed a much stronger function of the nanowire width than the Stokes peak.
We note that the frequency of the Stokes peak decreases with decreasing wire width more than seen in the model.
This can be a consequence of the approximate spin-wave dispersion relation used; the magnon wavevector $-q_m$ may not be as close to the bottom of the spin-wave band as in the model.
As a result, spin-wave quantization will lead to a reduction in the frequency of the mode with decreasing wire width. On reversing the field ($p=-1 \rightarrow p=+1$) the situation is reversed: the Stokes peak is now the higher frequency mode and its BLS spectra are more strongly affected by wire width.
These characteristics taken together are strong evidence that our model is capturing the essential physics. 

In summary, DMI in combination with the unusual nature of confined spin-wave modes in nanowires leads to a strong decrease in the frequency shift in light scattered by counter-propagating modes.
The strong frequency reduction is associated with mode quantization of quantized spin waves in the presence of DMI. 
This demonstrates that, in contrast to the BLS frequency shift in the longitudinal scattering geometry, which enables direct determination of the DMI~\cite{nembach2015linear}, the BLS frequency shift in the transverse scattering geometry is not directly related to the DMI.
This observation also raises the question of how such spin wave quantization affects other nanowire magnetic properties, such as skyrmions and domain wall dynamics in racetracks.
More generally, this physics is important for understanding spin waves in confined systems and characterizing antisymmetric exchange interactions in magnetic racetracks and other types of magnetic nanostructures that lack inversion symmetry.

This work was supported in part by the DARPA Topological Excitations in Electronics (TEE) program (grant
D18AP00009). A.D.K. also acknowledges support from NSF DMR-2105114. The nanostructures were realized at the Advanced Science Research Center NanoFabrication Facility of the Graduate Center at the City University of New York.

\section{Appendix}
\label{Sec:Appendix}
\subsection{BLS Results for different angles of incidence}
\label{Sec:Appendix_Angles}
\begin{figure}[ht!]
    \centering
    \includegraphics[width=0.4\textwidth]{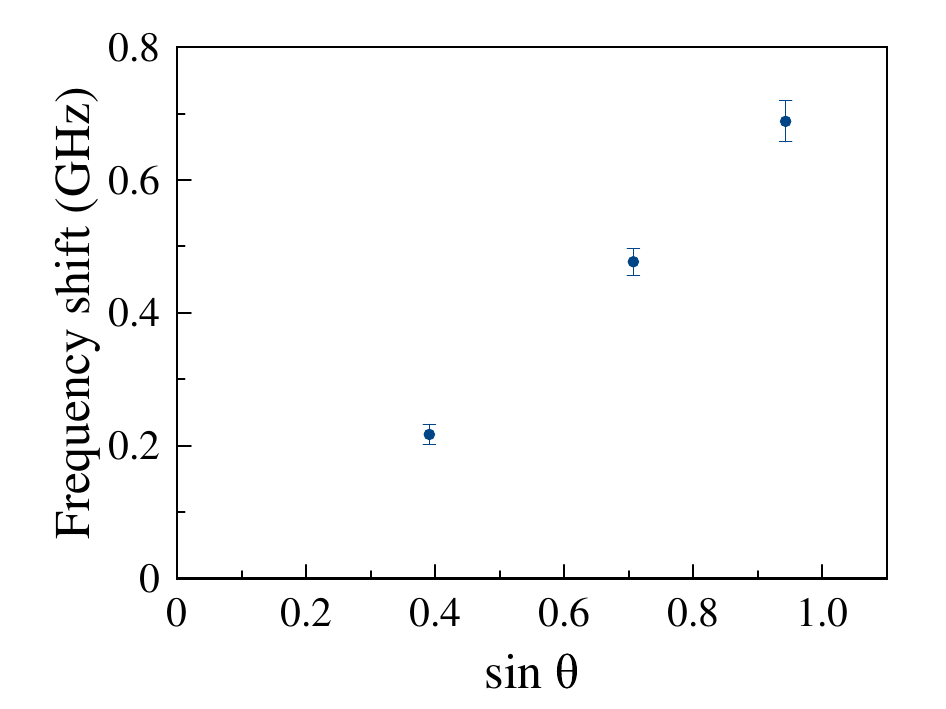}
    \caption{BLS frequency shift versus sine of the incident angle for the $\SI{400}{nm}$ nanowire array in the transverse scattering geometry.}
    \label{fig_s_400}
\end{figure}

We conducted BLS measurement in the transverse geometry for several different angles of incidence $\theta = 23^\circ, 45^\circ, 71^\circ$, to vary the in-plane wavevector $q_m$.
According to Eq.~1 in the main text, the frequency shift is proportional to the transferred wavevector $q_m$, and thus to $\sin \theta$.
In Fig.~\ref{fig_s_400}, we plot the frequency shift for the $\SI{400}{nm}$ nanowire array versus $\sin \theta$, which shows the expected linear relation.

\begin{figure}[ht!]
    \centering
    \includegraphics[width=0.4\textwidth]{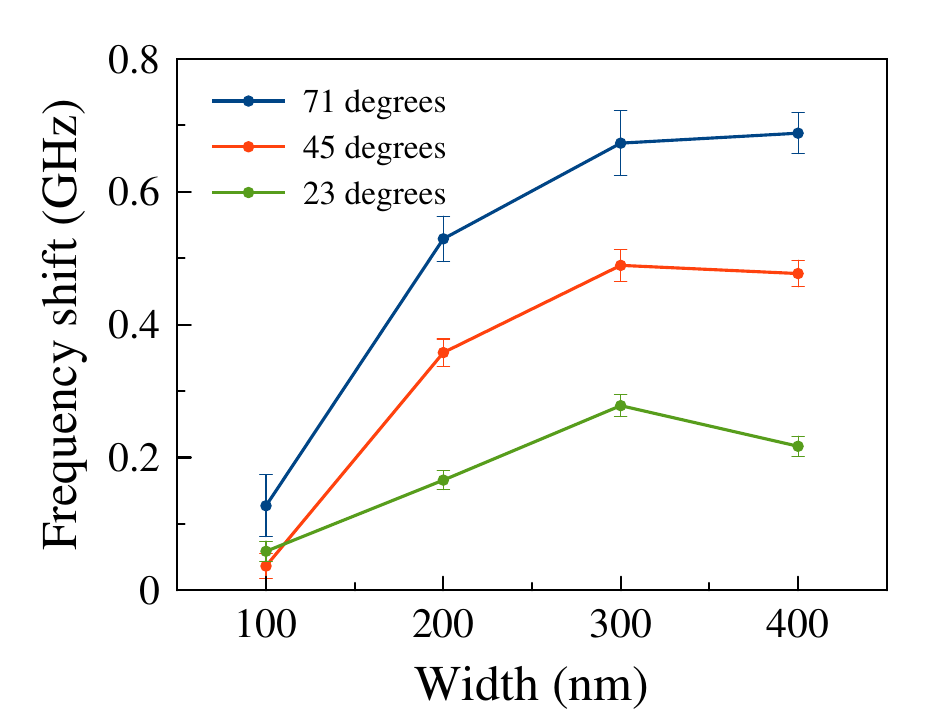}
    \caption{Nanowire width versus frequency shift for different incident angles in the transverse scattering geometry.}
    \label{fig_s_shfit}
\end{figure}

In Fig.~\ref{fig_s_shfit}, we plot the BLS frequency shift for nanowire arrays with different widths and different angles of incidence, again in the transverse scattering geometry.
Like the $\theta=45^\circ$ results discussed in the main text, a dramatic decrease in the frequency shift occurs when the nanowire width is less than $\SI{200}{nm}$.

\begin{figure}[ht!]
    \centering
    \includegraphics[width=0.4\textwidth]{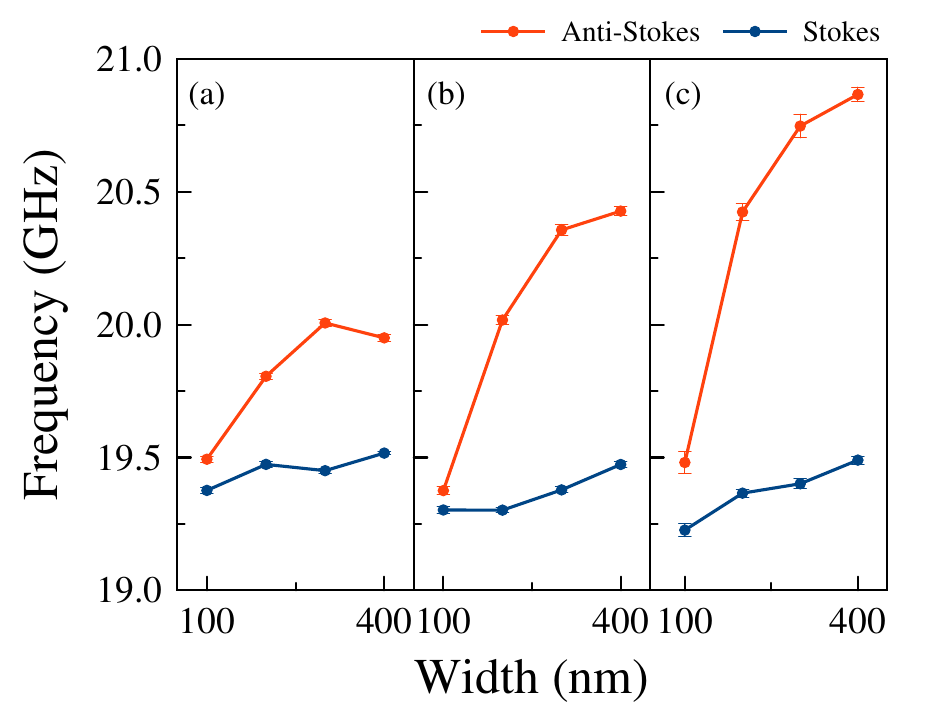}
    \caption{The Stokes and anti-Stokes peak frequencies for different nanowire widths for (a) $\theta = 23^\circ$, (b) $\theta = 45^\circ$, (c) $\theta = 71 ^\circ$ at a negative applied field, $p=-1$.}
    \label{fig_s_frequency}
\end{figure}

Lastly, in Fig.~\ref{fig_s_frequency} we plot the Stokes and anti-Stokes peak frequencies for different nanowire widths and angles of incidence.
As we showed in Fig.~4 in the main text, the major contribution to the decrease in the frequency shift comes from decreasing frequency of the anti-Stokes peak for this polarity of the applied field ($p=-1$).

\subsection{Ferromagnetic resonance spectroscopy}
\label{Sec:Appendix_FMR}
\begin{figure}[ht!]
    \centering
    \includegraphics[width=0.4\textwidth]{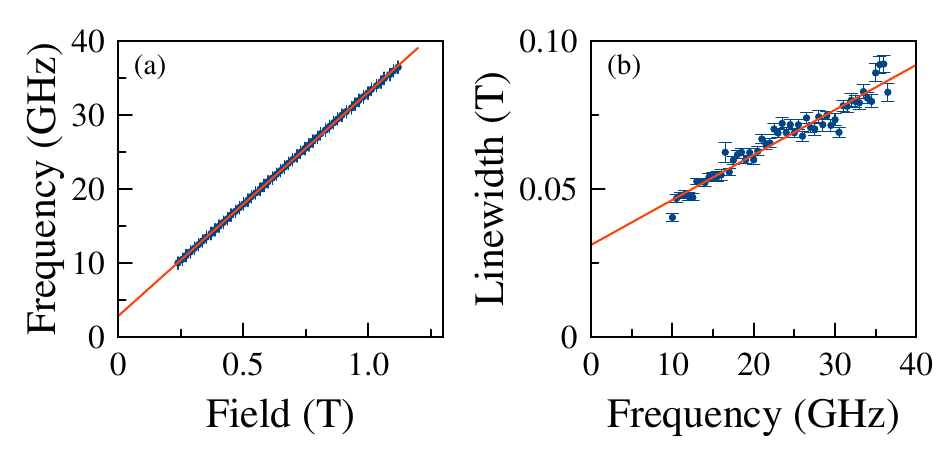}
    \caption{(a) FMR frequency versus resonance field in the field perpendicular geometry. The blue points are the experimental data and the orange line is a fit of the data to Eq.~\ref{eq_s_frequency}.
    (b) FMR linewidth versus frequency. The blue points are the data and orange line is a fit to Eq.~\ref{eq_s_linewidth}.}
    \label{fig_s_FMR}
\end{figure}

Ferromagnetic resonance (FMR) spectroscopy was conducted with a vector network analyzer (VNA).
The unpatterned film was diced and placed on a coplanar waveguide (CPW).
Polymethyl methacrylate (PMMA) was spun on the film's surface to prevent direct electrical contact to the CPW.
We applied a microwave signal to the CPW at a fixed frequency and swept the external magnetic field perpendicular to the film plane.
We then fit the complex transmission parameter $S_{21}$ to determine the resonance frequency and linewidth~\cite{nembach2011perpendicular}.
The out-of-plane resonance field and frequency follow the Kittel relation:
\begin{equation}
f = \frac{\mu_0 \gamma}{2 \pi} \rdbkt{H - M_\rm{eff}} \, ,
    \label{eq_s_frequency}
\end{equation}
where $\gamma$ is the gyromagnetic ratio, $\mu_0$ is the permeability of free space and $M_\rm{eff}$ is the effective magnetization. The linewidth versus frequency gives information on the damping:
\begin{equation}
    \mu_0 \Delta H = \frac{4 \pi \alpha f}{\gamma} + \mu_0 \Delta H_0 \, ,
    \label{eq_s_linewidth}
\end{equation}
where $\alpha$ is Gilbert damping constant and $\mu_0 \Delta H_0$ is the inhomogeneous linewidth broadening.
Figure~\ref{fig_s_FMR} shows the results with the fits to Eqs.~\ref{eq_s_frequency} and~\ref{eq_s_linewidth}.
We find $\gamma/2 \pi = \SI{30.3 \pm 0.1}{GHz/T}$, $M_\rm{eff} = \SI{-7.35 \pm 0.08 e4}{A/m}$ and $\alpha = \SI{0.0230 \pm 0.0008}{}$.

\subsection{SQUID magnetometry}
\begin{figure}[ht!]
    \centering
    \includegraphics[width=0.4\textwidth]{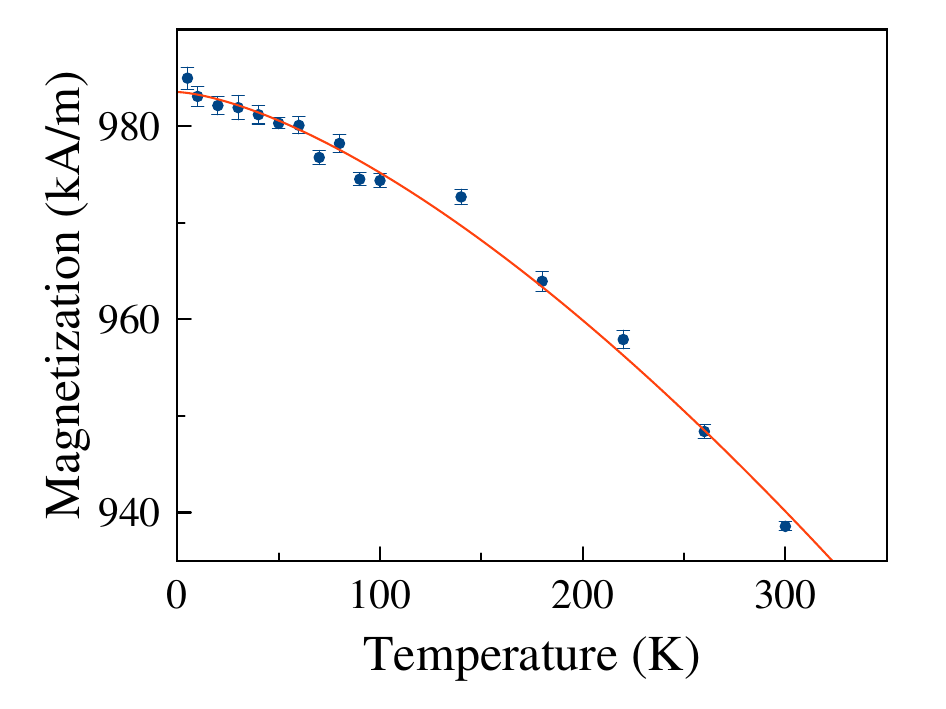}
    \caption{Saturation magnetization versus temperature. The blue points are the data and orange line is a fit to Eq.~\ref{Eq:Bloch}, which used to determine the exchange constant $A$.}
    \label{fig_s_SQUID}
\end{figure}

We used a SQUID magnetometer to measure the film's saturation magnetization. At room temperature, $M_\rm{s} = \SI{9.39 \pm 0.01 e5}{A/m}$.
In order to determine the exchange constant $A$, we measured the temperature dependence of the saturation magnetization $M_\rm{s}\rdbkt{T}$, as shown in Fig.~\ref{fig_s_SQUID}.
According to the Bloch $T^{3/2}$ law~\cite{vaz2008magnetism}, the temperature dependence follows the relation
\begin{equation}
    M_\rm{s}(T) = M_\rm{s}\rdbkt{0} \sqbkt{1 - \frac{g \mu_\rm{B} \eta}{M_\rm{s}\rdbkt{0}} \rdbkt{\frac{k_\rm{B} T}{D_\rm{spin}\rdbkt{0}}}^\frac{3}{2}} \, ,
    \label{Eq:Bloch}
\end{equation}
where $D_\rm{spin}\rdbkt{T}$ is the spin-wave stiffness, $g$ is Land\'e g-factor, $\mu_\rm{B}$ is the Bohr magneton and $\eta = 0.0587$ is a dimensionless constant that depends on the sample geometry.
In mean field theory~\cite{atxitia2007micromagnetic}, the temperature dependent spin-wave stiffness is given by
\begin{equation}
    D_\rm{spin}\rdbkt{T} = \frac{M_\rm{s}\rdbkt{T}}{M_\rm{s}\rdbkt{0}} D_\rm{spin}\rdbkt{0} \, .
\end{equation}
Then, we can calculate the exchange stiffness according to the equation
\begin{equation}
    A(T) = \frac{M_\rm{s}\rdbkt{T} D_\rm{spin}\rdbkt{T}}{2 g \mu_\rm{B}} \, ,
\end{equation}
and find $A = \SI{2.27 \pm 0.01 e-11}{J/m}$ at $\SI{300}{K}$. Further details on this analysis can be found in Ref.~\cite{nembach2015linear}.

\subsection{Width dependence of demagnetization factors}
\label{Sec:DM}
We did not include the width dependence of the nanowire's demagnetization factors into the analysis for several reasons.
First, the demagnetization factors change the frequency for the Stokes and anti-Stokes processes in the same manner and thus do not affect the frequency shift caused by DMI.
Secondly, only the frequency of the lowest mode ($n=0$) is effected, and the effect is negligible for all modes with higher indices (i.e., $n \geq 1$)~\cite{jorzick1999brillouin}.
Lastly, in our study, the Co thickness is only $\SI{1.8}{nm}$, which is much smaller than the width of the nanowires.
The width direction demagnetization factors calculated using a cuboid model are only 3.2\% for the $\SI{100}{nm}$ nanowires and 1.0\% for the $\SI{400}{nm}$ nanowires.
Therefore, ignoring the width direction demagnetization fields does not affect our analysis.

\subsection{Lifetime Broadening of BLS Spectra}\
\label{Sec:Appendix_Lifetime}
The lifetime of the spin-wave modes is limited by the magnetic damping.
To consider the spin-wave lifetime in our analysis and model, we broaden the model intensities $I_n$ in Eq.~7 of the main text by including the lifetime of the spin-wave modes.
First, the discretized intensity can be written as 
\begin{equation}
    I(f) = \sum_{m=0}^\infty I_n \delta \rdbkt{f - f_n} \, ,
\end{equation}
where $\delta \rdbkt{f - f_n}$ is the Dirac delta function.
We then convolute the delta function with a Lorentzian function of the form:
\begin{equation}
    L_n(f) = \frac{1}{\rdbkt{f - f_n}^2 + \rdbkt{\frac{\Delta f_n}{2}}^2} \,.
\end{equation}
The mode lifetime is set by the Gilbert damping constant, Eq.~\ref{eq_s_linewidth} and thus $\Delta f_n = \Delta H \left. \dv{f}{H} \right|_{f = f_n} $, i.e., we convert the linewidth in field to a frequency linewidth.
The curves shown in Fig.~3(c-f) in the main text, before normalization, are then given by
\begin{equation}
    I(f) = \sum_{n=0}^\infty \frac{I_n}{\rdbkt{f - f_n}^2 + \rdbkt{\frac{1}{2}  \Delta H \left.\dv{f}{H}\right|_{f = f_n}}^2} \, ,
\end{equation}
where $\mu_0 \Delta H$ is the linewidth taken from FMR experiment.

\bibliographystyle{apsrev4-1}

\end{document}